\documentclass[draft]{agujournal}

\usepackage{soul}

\journalname{JGR-Planets}

\begin{document}

%
%

\title{Gravity and Zonal Flows of Giant Planets: \\From the Euler Equation to the Thermal Wind Equation}

%
%




\authors{Hao Cao\affil{1},
 David J. Stevenson\affil{1}}

\affiliation{1}{Division of Geological and Planetary Sciences, California Institute of Technology, Pasadena, CA 91125}






\correspondingauthor{Hao Cao}{haocao@caltech.edu}




\begin{keypoints}
\item We present a critical examination of the applicability of the thermal wind equation to calculate the wind induced gravity moments.
\item Full solution of wind-induced gravity moments from the Bessel method and Concentric Maclaurin Spheroid method agree very well.
\item Solution of wind-induced gravity moments from the nonspherical thermal wind equation agrees well with the full solution.
\end{keypoints}

%
%


\begin{abstract}
Any nonspherical distribution of density inside planets and stars gives rise to a non-spherical external gravity and change of shape. If part or all of the observed zonal flows at the cloud deck of Jupiter and Saturn represent deep interior dynamics, then the density perturbations associated with the deep zonal flows could generate gravitational signals detectable by the \textit{Juno} mission and the \textit{Cassini Grand Finale}. Here we present a critical examination of the applicability of the thermal wind equation to calculate the wind induced gravity moments. Our analysis shows that wind induced gravity moments calculated from TWE are in overall agreement with the full solution to the Euler equation. However, the accuracy of individual high-degree moments calculated from TWE depends crucially on retaining the nonsphericity of the background density and gravity. Only when the background nonsphericity of the planet is taken into account, does the TWE make accurate enough prediction (with a few tens of percent errors) for individual high-degree gravity moments associated with deep zonal flows. Since the TWE is derived from the curl of the Euler equation and is a local relation, it necessarily says nothing about any density perturbations that contribute irrotational terms to the Euler equation and that have a non-local origin. However, the predicted corrections from these density contributions to the low harmonic degree gravity moments are not discernible from insignificant changes in interior models while the corrections at high harmonic degree are very small, a few percent or less. 
\end{abstract}

%
%

%


%
%
%
%

\section{Introduction}

The interior structures and dynamics of the solar system giant planets remain elusive after decades of observational, experimental and theoretical studies \citep[cf.][and reference therein]{Stevenson1982, HBL2002, Guillot2005, GG2014}. For example, whether present-day Jupiter and Saturn have well-defined cores remains an open question; the total enrichment of heavy elements inside Jupiter and Saturn are not well constrained; the structural and dynamical consequences of the likely on-going sedimentation of helium and neon inside Jupiter and Saturn have not been fully worked out \citep[cf.][]{Stevenson1977, FH2003, Nettelmann2015}. 

One lasting debate concerns the nature of the observed east-west zonal flows on the cloud layers of giant planets with amplitude on the order of 100 $m/s$: no consensus has been reached upon whether these zonal winds represent shallow atmospheric dynamics or deep interior dynamics \citep[e.g.][]{VS2005, LGS2008, Jones2009, Kaspi2009, LS2010, GWA2013}. The forward fluid dynamics and magnetohydrodynamics (MHD) problem about the nature of giant planet zonal flows may be hard to settle given the complexity of the system and the extreme parameters involved. However, an observational fact about the depth of the zonal flows of Jupiter and Saturn will likely be established given the gravity and magnetic experiments from the \textit{Juno} mission \citep[]{Bolton2010} and the upcoming \textit{Cassini Grand Finale} \citep[]{Spilker2014}. In this paper, we focus on the gravity field. 

The physical principle of the gravitational sounding of giant planet zonal flows is not complicated: zonal flows will induce local and non-local density perturbations, as well as global shape change of the planet, all of which will contribute to perturbations to the external non-spherical gravity field. Apart from observational issues such as data coverage, the analysis of the actual gravity measurement is further complicated by the fact that the background external non-spherical gravity field caused by the background uniform rotation is not known a priori. Even if one would like to analyze the problem in real space (e.g. directly assess the gravity field $\mathbf{g}$ rather than the gravity moments $J_n$), one is still forced to analyze the truncated gravity field associated with high-degree gravity moments only (e.g. $\Delta \mathbf{g}$ associated with $J_n$, $n \ge 12$ for Jupiter and Saturn), in order to retrieve information about the contribution of differential rotation to the gravity field. Since the gravity measurements at Jupiter and Saturn will not be sensitive to an infinite series of high-degree gravity moments due to the geometric decay, the accuracy of the individual high-degree gravity moments associated with zonal flows from a forward model is an important issue.

The thermal wind equation (TWE) under the anelastic approximation can be used to calculate the gradient of local density perturbations $\nabla \rho'$ associated with zonal flows, when the zonal flows are much slower than the background rotation. The measured differential rotation on the surface of Jupiter and Saturn is small compare to the background planetary rotation. In terms of the Rossby number $Ro = u/\Omega_0R_p$ (u is the velocity measured in the corotating frame, $\Omega_0$ is the background rotation rate, $R_p$ is the planetary radius), $Ro$ at Jupiter is smaller than 0.01, and Ro at Saturn is smaller than 0.05. However, the applicability of TWE to further calculate the individual gravity moments for an oblate planet is not guaranteed a priori, given the non-local nature of the gravity moments. This applicability has been actively debated in the recent literature \citep[cf.][]{Kaspi2010, Kong2014, ZKS2015, Kaspi2016, Galanti2017}.

In this paper, we present a critical evaluation of the applicability of the thermal wind equation to calculate the zonal flow gravity moments. The gravity moments associated with deep zonal flows calculated from two versions of the thermal wind equation are compared to the full solution to the Euler equation obtained from the Bessel method \citep[]{Hubbard1975, Wisdom1996, Hubbard1999} and the Concentric Maclaurin Spheroid (CMS) method \citep[]{Hubbard2013}. A suite of barotropic wind profiles are evaluated. We found that wind induced gravity moments calculated from TWE are in overall agreement with the full solution to the Euler equation. However, only when the non-sphericity of the background density and effective gravity is taken into account, is the individual high-degree gravity moment calculated from the thermal wind equation a good approximation to the full solution. Our analysis thus suggests that, when analyzing the zonal flow gravity moments of Jupiter and Saturn using the thermal wind equation, the non-sphericity of the background state should be retained.  

The paper is organized as following, section 2 introduces the definition and properties of gravity moments, section 3 presents a detailed comparison of the Euler equation and the thermal wind equation, section 4 presents the gravity moments of a uniformly rotating planet with polytrope of index unity calculated from the Euler equation using the Bessel method and the CMS method, section 5 presents the gravity moments of a differentially rotating planet with polytrope of index unity calculated from the Euler equation, the thermal wind equation with spherical background state, and the thermal wind equation with non-spherical background state, section 6 presents an analysis of what the thermal wind equation misses, section 7 summarizes the results and discusses the implications for analyzing the gravity measurements of the \textit{Juno} mission and the \textit{Cassini Grand Finale}. 

\section{Definition and Properties of the Gravity Moments}

The axisymmetric gravity moments $J_n$ are determined by the planetary interior density distribution through
\begin{equation}
J_n=-\frac{1}{Ma^n} \int_{\mathbf{R}^3} \rho(\mathbf{r}) r^n P_n(cos\theta) d^3\mathbf{r},
\label{eqn:JnDef}
\end{equation}
in which $M$ is the mass of the planet, $a$ is a reference radius usually chosen to be the measured equatorial radius of the planet, $r$ is the spherical-radial distance to the center of mass of the planet, $P_n$ are the Legendre polynomials of degree $n$, $\theta$ is the co-latitude measured from the spin-axis and the integration is over the entire volume of the planet. 

It should be immediately realized that 1) if the density distribution is spherically symmetric, $\partial \rho/\partial \theta=0$, all $J_n$ with $n \ge 1$ would be zero; 2) if the density distribution is equatorially symmetric, $\rho(r,\theta)=\rho(r,\pi-\theta)$, all odd-degree $J_n$ would be zero. 


If mass, equatorial radius, rotation rate and the gravity moments are the only measurements we have about a planetary body, the interpretation of gravity moments depend on extra assumptions/knowledge and forward modeling of the density distributions inside the planet. The extra assumptions/knowledge include the composition, temperature, and the equation of state (EOS) of the relevant material. The appropriate forward model for the density distribution inside a fluid planet, for a given EOS, is nothing but the appropriate governing equations of fluid dynamics. In the inviscid limit, this set of governing equation is the Euler equation. Even though the dynamics being considered are usually simple (e.g. uniform rotation or differential rotation on cylinders only), the Euler equations for this particular application are not easy to solve due to the fact that we are dealing with self-gravity. The problem is non-local: gravity at any local position depends on the density distribution over the entire planet. Mathematically, one needs to deal with integro-differetial equations in general. 

\section{From the Euler Equation to the Thermal Wind Equation}


\subsection{The Euler Equation}

The structure and dynamics of a self-gravitating fluid body in steady-state must be in force-balance. This force-balance in the inviscid limit is described by the steady-state Euler equation. In an inertial frame, the Euler equation reads
\begin{equation}
(\mathbf{u} \cdot \nabla) \mathbf{u}=-\frac{\nabla P}{\rho} -\nabla V_g,
\label{eqn:Euler}
\end{equation}
in which $\mathbf{u}$ is the velocity in the inertial frame, $P$ is the pressure, $\rho$ is the density, and $V_g$ is
the gravitational potential, the negative gradient of which is the gravitational acceleration:
\begin{equation}
\mathbf{g} = -\nabla V_g. 
\label{eqn:GravPot}
\end{equation}

Considering self-gravity only, the gravitational potential is determined by the global density distribution through  
\begin{equation}
V_g (\mathbf{r})= - \int_{\mathbf{R}^3} \frac{G}{|\mathbf{r}-\mathbf{r'}|}\rho(\mathbf{r'})d^3\mathbf{r'},
\label{eqn:GravPotRho}
\end{equation}
in which $G$ is the gravitational constant, and the integration is over the entire domain of the planet. 


Under the barotropic assumption (density depends on pressure only) and a velocity field that does not violate the barotropic assumption, the density distribution can be determined entirely from the Euler equation (\ref{eqn:Euler}) given the total mass and the specific equation of state. If the fluid is baroclinic, an additional equation governing the evolution of temperature or entropy is needed to determine the density distribution.  

\subsection{The Euler Equation in an Inertial Frame for a Uniformly Rotating Planet}

For a uniformly rotating planet, the velocity in the inertial frame reads
\begin{equation}
\mathbf{u_0} =\Omega_0 s \hat{\phi}= \Omega_0 r \sin \theta \hat{\phi},
\label{eqn:u0}
\end{equation}
here $\Omega_0$ is the constant angular velocity and $s$ is the cylindrical radial distance from the spin axis ($s=r \sin \theta$).

It can be easily shown that
\begin{equation}
(\mathbf{u_0} \cdot \nabla) \mathbf{u_0} = \nabla Q_0
\label{eqn:u0Q0}
\end{equation}
in which $Q_0$ is the familiar centrifugal potential
\begin{equation}
Q_0=-\int_0^s \Omega_0^2 s' ds' = - \frac{\Omega_0^2 s^2}{2}.
\label{eqn:Q0} 
\end{equation}

The Euler equation now reads
\begin{equation}
\nabla P = -\rho \nabla (V_g + Q_0) = -\rho \nabla U, 
\label{eqn:Euler0} 
\end{equation}
where $U$ is the effective potential defined as $U=V_g+Q_0$.

When coupled with a specific equation of state (EOS), the solution of the above Euler equation yields the shape and internal density distribution of a uniformly rotating planet. It is appropriate to denote the properties satisfying equation (\ref{eqn:Euler0}) as the background properties with a subscript $0$, so equation (\ref{eqn:Euler0}) now reads
\begin{equation}
\nabla P_0=-\rho_0 \nabla (V_{g_0} + Q_0) = -\rho_0 \nabla U_0.
\label{eqn:Euler00}
\end{equation}


\subsection{The Euler Equation for a Planet with Differential Rotation and the Thermal Wind Equation}

Now consider a planet with differential rotation, the velocity in the inertial frame reads 
\begin{equation}
\mathbf{u_1} = [\Omega_0 + \Omega'(s, z)] s \hat{\phi} = \Omega_1 s \hat{\phi},
\label{eqn:u1}
\end{equation}
here $\Omega'$ is the angular velocity of the differential rotation, $\Omega_1$ is the total angular velocity measured in the inertial frame, while
\begin{equation}
\mathbf{u'} = \mathbf{u_1} - \mathbf{u_0} = \Omega'(s, z) s \hat{\phi}
\label{eqn:uprim}
\end{equation}
is the zonal velocity measured in the non-inertial frame rotating at angular velocity $\Omega_0$.

If the angular velocity of the differential rotation depends only on the cylindrical radius ($\partial \Omega'/\partial z=0$), it can be shown that 
\begin{equation}
(\mathbf{u_1} \cdot \nabla) \mathbf{u_1} = \nabla Q, 
\end{equation}
here $Q$ is the generalized centrifugal potential
\begin{equation}
Q= -\int_0^s \Omega_1^2 s' ds' = - \int_0^s [\Omega_0 + \Omega'(s')]^2 s' ds'.
\label{eqn:Q1}
\end{equation}

For an arbitrary flow, it can be shown that,
\begin{equation}
(\mathbf{u_1} \cdot \nabla) \mathbf{u_1} = \nabla Q_0 + 2 \Omega_0 \hat{z} \times \mathbf{u'} + (\mathbf{u'} \cdot \nabla) \mathbf{u'},
\label{eqn:u1du1}
\end{equation}
it should be recognized that $-\nabla Q_0$ and $-2 \Omega_0 \hat{z} \times \mathbf{u'}$ are simply the centrifugal acceleration and the Coriolis acceleration in the rotating frame with angular velocity $\Omega_0$.

Substitute equation (\ref{eqn:u1du1}) and equation (\ref{eqn:Euler00}) into the Euler equation (\ref{eqn:Euler}), and write the density and pressure as the sum of the background and the perturbation
\begin{equation}
\rho_1 = \rho_0 + \rho'
\end{equation}
\begin{equation}
P_1 =P_0 +P',
\end{equation}
we get
\begin{subequations}
\begin{align}
2\rho_0 & \Omega_0 \hat{z} \times \mathbf{u'}  + \rho_0(\mathbf{u'} \cdot \nabla) \mathbf{u'} + \rho_0 \nabla V_{g'} = \tag{17a}\label{eq:EulerRot_1}\\ 
 & \qquad{} \quad{} -\rho' \nabla V_{g_0} -  \rho' \nabla Q_0 - \nabla P'   \tag{17b}\label{eq:EulerRot_2} \\
 & - 2 \rho' \Omega_0 \hat{z} \times \mathbf{u'} - \rho' (\mathbf{u'} \cdot \nabla) \mathbf{u'} -\rho' \nabla V_{g'}, \tag{17c}\label{eq:EulerRot_3}
\end{align}
\end{subequations}
where $V_{g'}$ is the gravitational potential associated with the density perturbations $\rho'$
\begin{equation}
V_{g'}(\mathbf{r})= - \int_{\mathbf{R}^3} \frac{G}{|\mathbf{r}-\mathbf{r'}|}\rho'(\mathbf{r'})d^3r'.
\end{equation}

An order of magnitude analysis yields the first estimate about the relative importance of each of the terms in equation (17). 1) The ratio between the advection term associated with the zonal flows $\rho_0(\mathbf{u'} \cdot \nabla) \mathbf{u'}$ and the Coriolis term $2\rho_0 \Omega_0 \hat{z} \times \mathbf{u'}$ is on the order of the Rossby number: $\sim$ 1\% for Jupiter and $\sim$ 5\% for Saturn. 2) The gravity anomaly associated with the density perturbation, $\nabla V_{g'}$, is smaller than the background gravity by a factor of $\rho'/\rho$. \citet[]{ZKS2015} pointed out that $\rho_0 \nabla V_{g'}$ could be comparable to $\rho' \nabla V_{g_0}$. 3) The gradient of the pressure perturbation $\nabla P'$ is likely comparable to $\rho' \nabla U_0$. This can be shown easily for a polytropic equation of state $P=K\rho^{(1+1/n)}$. 
The perturbative pressure can now be expressed as a function of the density
\begin{subequations}
\begin{align}
P' & = P-P_0 \tag{19a}\label{eq:Pp_1}\\ 
    & =K(\rho_0+\rho')^{(1+1/n)}-K\rho_0^{(1+1/n)}. \tag{19b}\label{eq:Pp_1}
\end{align}
\end{subequations}
Assuming $\rho'/\rho_0 \ll 1$, one can Taylor expand the above equation. Retaining the first order term only, we get 
\begin{equation}
P'=(1+\frac{1}{n})K\rho_0^{1/n}\rho'.
\label{eqn:PpT}
\end{equation}
Taking the gradient of $P'$, and making use of the hydrostatic balance of the background state, we get
\begin{equation}
\nabla P' = -\frac{1}{n}\rho' \nabla U_0 + (1+\frac{1}{n})K\rho_0^{1/n}\nabla \rho'. \label{eq:GradPp_1}
\end{equation}
(Note, the second term in the RHS of $(21)$ can be comparable to the first term $\frac{1}{n}\rho' \nabla U_0$ given the likely small characteristic scale of $\rho'$.) And 4) the ratio of each of the last three terms in equation (17) to its corresponding LHS term is $|\rho'/\rho_0|$. One cautionary note about the value of $|\rho'/\rho_0|$. As we will see, although it is true that $|\rho'/\rho_0|$ is smaller than one in the bulk interior of the planet, this is not true for regions very near the surface.

Taking the curl of the equation ($17$), retaining only the first term on the left hand side (LHS) and the first two terms on the right hand side (RHS), and making use of the mass continuity equation under the anelastic approximation, we arrive at the generic thermal wind equation (TWE)
\begin{equation}
(2\mathbf{\Omega_0} \cdot \nabla)(\rho_0 \mathbf{u})=-\nabla \rho' \times \mathbf{g_{eff}},
\label{eqn:eTWE}
\end{equation}
where the background effective gravity $\mathbf{g_{eff}}$ is
\begin{equation}
\mathbf{g_{eff}}=-\nabla U_0= - \nabla (V_{g0} + Q_0).
\label{eqn:EffGravRho} 
\end{equation}

Note, first, that since the curl has been taken, information has been lost. Specifically, there are solutions to the curl free part of equation which can contribute to density perturbations. In addition, there will be density perturbations with non-local origin because of 1) the gravity resulting from the local density anomalies that are required by the TWE and 2) the global shape change associated with the net angular moment of the zonal flows. The non-local density perturbations associated with the gravity anomaly resulting from the local density perturbations required by the TWE was recognized by \citet[]{ZKS2015}. However, as we will show in section 6, this type of non-local density perturbation contributes very little to the high-degree gravity moments.

One further simplification to the generic thermal wind equation (\ref{eqn:eTWE}) usually adopted in estimating the zonal flow gravity field is to assume that the background effective gravity is spherically symmetric \citep[e.g.][]{Kaspi2010, Liu2013}. One argument for this simplification is the uniqueness of $J_n$ calculated under this assumption despite a non-uniqueness in the density perturbations calculated from the TWE. However, we will show that the mathematical uniqueness gained from this assumption \textbf{is} not worth the physical relevance being sacrificed. And the mathematical non-uniqueness in the density perturbations from the TWE can be treated through physically reasonable assumptions.  

\section{Gravity Moments of a Uniformly Rotating Planet with Polytrope of Index Unity}

To compare the density perturbations and gravity moments associated with the deep zonal flows calculated from the thermal wind equation to the full solution of the Euler equation, we first solve the Euler equation for a uniformly rotating planet. Here, we adopt a polytropic equation of state
\begin{equation}
P=K\rho^{(1+1/n)},
\label{eqn:PolyT} 
\end{equation}
in which K is a constant, and the polytropic index $n$ is set to 1. A polytrope of index unity not only is a reasonable approximation to the adiabatic equation of state under Jupiter conditions but also makes the Euler equation easier to deal with since $\nabla P/\rho$ now reduces to $2K\nabla \rho$. The divergence of the Euler equation (\ref{eqn:Euler00}) yields
\begin{equation}
\nabla^2 \rho_0 +\frac{2\pi G}{K} \rho_0 = \nabla^2 (\frac{\Omega_0^2s^2}{4K}),
\label{eqn:rhoPoly1}
\end{equation}
which governs the background density distribution subject to the boundary condition that the outer boundary defined by $\rho_0(r,\theta)=0$ is also an equipotential surface 
\begin{equation}
U_0(\rho_0(r, \theta)=0)=const.
\end{equation}

As shown by \citet[]{Hubbard1975}, \citet[]{Wisdom1996}, and \citet[]{Hubbard1999}, the general solution to the above equation takes the form
\begin{equation}
\rho=\rho_P+\sum_{n=0}^{n_{max}}A_n j_n(kr) P_n(\cos \theta),
\label{eqn:RhoPoly1} 
\end{equation}
where $\rho_P= \Omega_0^2/2 \pi G$, $j_n$ are the spherical Bessel functions of the first of kind of degree $n$, $k=\sqrt{2\pi G/K}$, and $P_n$ are the Legendre polynomials of degree $n$, and $A_n$ are the coefficients to be determined by the boundary conditions as well as the total mass. 

The existence of the analytical form of the general solution enables a non-perturbative approach to this problem \citep[]{Hubbard1975, Wisdom1996, Hubbard1999}. Under this circumstance, there is no need to define the level surfaces and solve for the figure equations explicitly. 
Here we point out a few key aspects of this non-perturbative approach which we call the Bessel method following \citet[]{Wisdom1996} and \citet[]{WH2016}: 1) the coefficients $A_n$ are the only variables that need to be solved explicitly, both the outer boundary shape, which defines the solution domain, and the internal density distribution are uniquely determined by $A_n$; 2) the outer boundary is not constrained to be an exact ellipsoid of revolution, and the resulting outer boundary indeed differs from an exact ellipsoid of revolution; 3) the traditional geophysical expansion of the external gravitational potential $U_0$ is used to calculate the potential at the outer boundary, given that its convergence under Jupiter and Saturn like surface distortions has been shown by \citet{Hubbard2014}. 

The Concentric Maclaurin Spheroid (CMS) method by \citet[]{Hubbard2013} is a non-perturbative method capable of solving for the equilibrium internal density distribution of a rotating planet with an arbitrary equation of state. Interested readers should refer to \citet[]{Hubbard2013} for the details of the method. Here we summarize a few key aspects of the CMS method: 1) the modeled planet is discretized into a finite number of concentric constant-density spheroids; 2) the shape of each constant-density spheroid is found iteratively via requiring it to be an equipotential surface; 3) the gravitational potential at each level surface is the sum of the contributions from every spheroid; 4) the density of each constant-density spheroid needs to be adjusted iteratively to match the prescribed equation of state and the fixed total mass of the planet. 

\begin{table}[t!]
\setlength{\tabcolsep}{4pt}
\renewcommand{\arraystretch}{1.0}
\caption{Gravity Moments of a Uniformly Rotating Polytrope of Index Unity Planet from Different Methods.\label{tab1}}
\centering
\begin{tabular}{ c | r r r r } 
\hline
\hline
 &  & Polytrope Model$^{b}$ & Polytrope Model$^{b}$ & Fractional \\
 &  Jupiter$^{a}$  & Bessel Method & CMS Method (521) &  Difference \\
\hline
Mass [$kg$]& $1.89819 \times 10^{27}$ & $1.8983 \times 10^{27}$ & $1.8983 \times 10^{27}$ \\ 
Rotation Period [$hrs$]  & $9.925$ & $9.925$  & $9.925$\\
Equatorial Radius [$km$] & $71492$ & $71418.75$ & $71407.15$ & $1.62\times10^{-4}$\\ 
Polar Radius [$km$] & $66854$ & $66875.28$ & $66868.54$ & $1.01 \times 10^{-4}$  \\ 
$J_2 \times 10^{-6}$ & $14696.43$ & $13948.95$ & $13953.32$ & $3.13\times10^{-4}$ \\
$J_4 \times 10^{-6}$ & $-587.14$ & $-528.81$  & $-529.14$ & $6.18 \times 10^{-4}$\\
$J_6 \times 10^{-6}$ & $34.25$ & $29.86$ & $29.89$ & $9.15 \times 10^{-4}$\\
$J_8$ &  & $-2.108 \times 10^{-6}$ & $-2.110 \times 10^{-6}$ & $1.20 \times 10^{-3}$ \\
$J_{10}$ &  & $1.716 \times 10^{-7}$ & $1.718 \times 10^{-7}$ & $1.48 \times 10^{-3}$ \\
$J_{12}$ &  & $-1.541 \times 10^{-8}$ & $-1.544 \times 10^{-8}$ & $1.76 \times 10^{-3}$ \\
$J_{14}$ &  & $1.488 \times 10^{-9}$ & $1.491 \times 10^{-9}$ & $2.02 \times 10^{-3}$ \\
$J_{16}$ &  & $-1.517 \times 10^{-10}$ & $-1.520 \times 10^{-10}$ & $2.28 \times 10^{-3}$ \\
$J_{18}$ &  & $1.614 \times 10^{-11}$ & $1.618 \times 10^{-11} $ & $2.54 \times 10^{-3}$\\
$J_{20}$ &  & $-1.777 \times 10^{-12}$ & $-1.782 \times 10^{-12}$ & $2.78 \times 10^{-3}$ \\
\hline
\multicolumn{5}{l}{$^{a}$ The physical values are from \citet[]{IAU2009} \& \citet[]{JUP230} with $G=6.67408 \times 10^{-11} m^3 kg^{-1} s^{-2}$.}\\
\multicolumn{5}{l}{$^{b}$ $P=K\rho^2$ with $K=2 \times 10^5$ [$m^5 kg^{-1} s^{-2}$].}
\end{tabular}
\end{table}


\begin{figure}[h!]
 \centering
     \includegraphics[width=\textwidth]{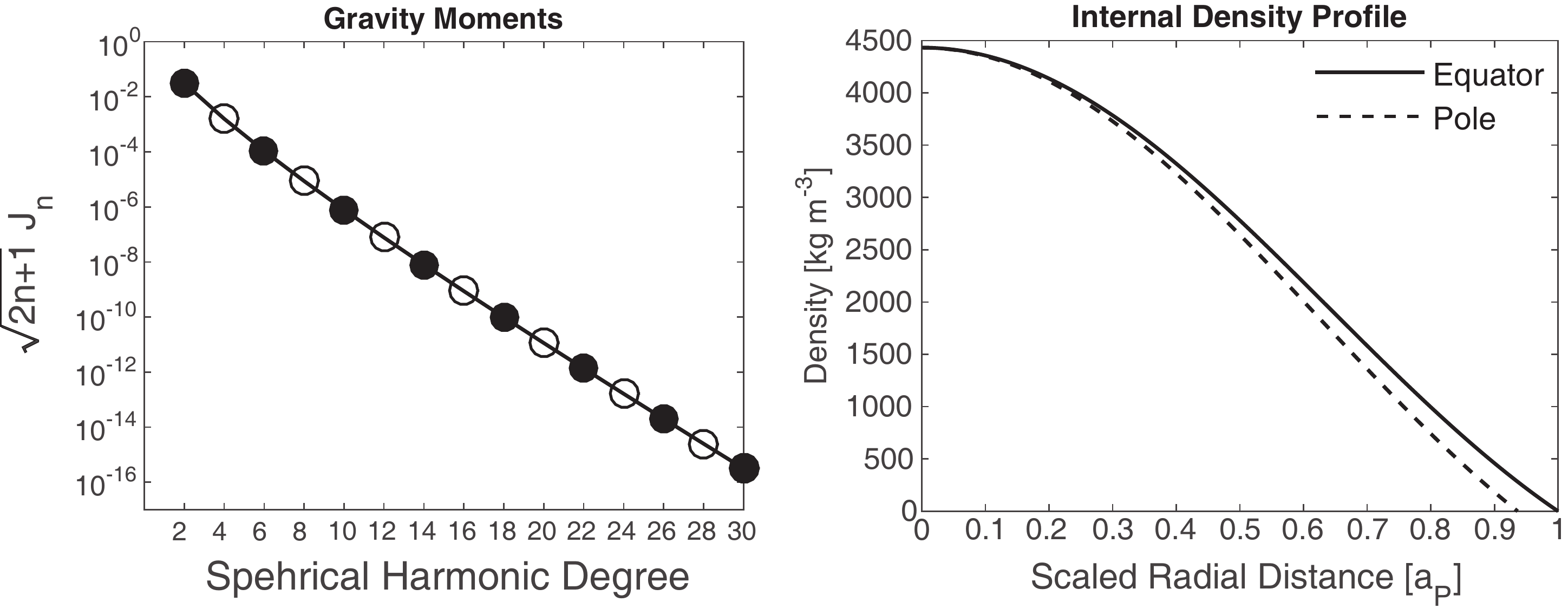}
 \caption{Gravity moments $J_n$ and internal density distribution of a uniformly rotating planet with fixed mass, rotation rate, and a polytropic equation of state of index unity with $K=2\times 10^5$  [$m^5 kg^{-1} s^{-2}$]. For $J_n$, filled (open) symbols represent positive (negative) values. For $J_n$, solutions to the full Euler equation from the Bessel method and the Concentric Maclaurin Spheroid method (with 521 spheroids) are shown here. Due to the excellent agreement between the two methods, the two solutions appear indistinguishable on this plot.}
\end{figure}

Fig. 1 shows the gravity moments and the internal density distribution from these two methods with $K= 2 \times 10^5$ [$m^5 kg^{-1} s^{-2}$]. The total mass and the background rotation period have been fixed to $1.8983 \times 10^{27} kg$ and 9.925 hours respectively, very close to the measured value of Jupiter. Tab. 1 compares solutions to the Euler equation (the equatorial radius, the polar radius, and the gravity moments up to degree 20) from the Bessel method and the CMS method for a uniformly rotating planet with a polytrope of index unity. It can be seen from Tab. 1 that the fractional differences of the gravity moments, defined as $|J_n(\text{Bessel})-J_n(\text{CMS})|/|J_n(\text{Bessel})|$, are on the order of $3 \times 10^{-4} \sim 3 \times 10^{-3}$. This agrees with the \citet[]{WH2016} assessment that the discretization error for a 512-spheroid CMS is on the order of $1 \times 10^{-4} \sim 1 \times 10^{-3}$. The observed values of Jupiter are listed as well. It can be seen that the first three gravity moments of this model planet are reasonably close to those measured at Jupiter. 

The right panel of Fig. 2 shows the effective ellipticity of the equipotential surface as a function of the equatorial radius of the equipotential surface. It can be seen that the effective ellipticity decreases from $\sim$ 0.35 near the surface of the planet to $\sim$ 0.29 near the center of the planet, and this shape change occurs mostly in the outer part of the planet. The effective ellipticity is defined as $e(r)=\sqrt{1-r_b^2/r_a^2}$, where $r_a$ and $r_b$ are the equatorial radius and the polar radius of level surfaces. We call this effective ellipticity due to the fact that the level surface are not exact ellipsoids of revolution. The left panel of Fig. 2 shows the deviation of the outer boundary equipotential surface and mid-shell equipotential surface from an exact ellipsoid of revolution with the equatorial radius and polar radius fixed to the corresponding values of the equipotential surface. The deviation is dominated by $\sin^2 2\theta$ with an amplitude $\sim 5 \times 10^{-4}$ at the outer boundary, and thus corresponds to the second order correction in the standard expansion of level surface in terms of the effective ellipticity (e.g. equation 30.3 in \citealt{ZT1978}). We noticed that some published solutions of this problem are based on the assumption that the outer boundary shape is an exact ellipsoid of revolution \citep[e.g.][]{Kong2013, Kong2015}.

\begin{figure}[h!]
 \centering
     \includegraphics[width=\textwidth]{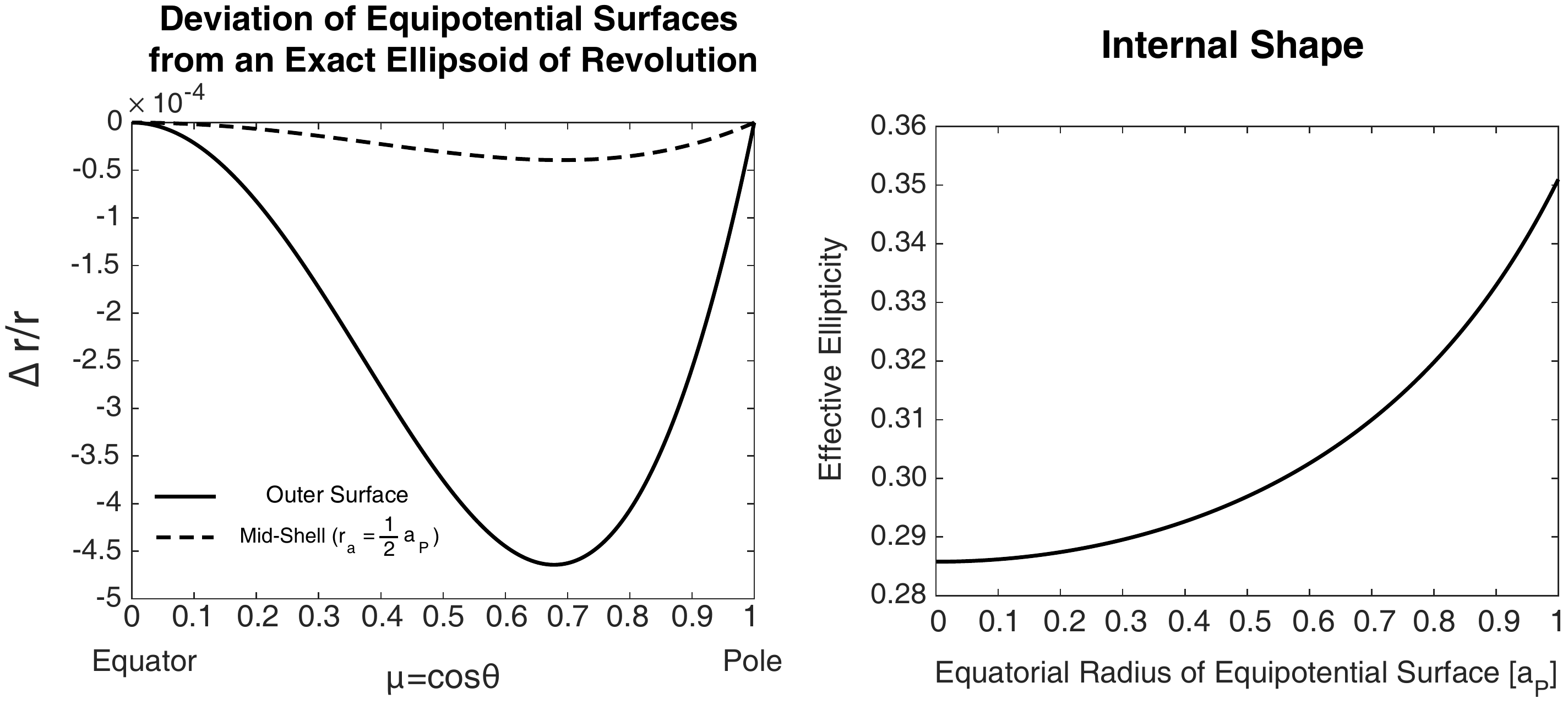}
 \caption{Deviation of equipotential surfaces from that of an exact ellipsoid of revolution with the equatorial radius and the polar radius fixed to the corresponding values of the equipotential surface, and the internal shape of equipotential surfaces measured by the effective ellipticity $e$=$\sqrt{1-r_b^2/r_a^2}$. The deviation of the outer boundary from an exact ellipsoid of revolution shows a dominant component of $\sin^2 2\theta$ with amplitude $\sim 5 \times 10^{-4}$. This corresponds to the second order correction in the standard expansion of level surface in terms of the effective eccentricity (e.g. equation 30.3 in \citealt{ZT1978}).}
\end{figure}

\section{Gravity Moments of a Differentially Rotating Planet with Polytrope of Index Unity}

We now turn to a planet with differential rotation. The total mass, background rotation rate, and the equation of the state are taken to be the same as those in the study of a uniformly rotating planet. Note that the equatorial radius and the shape of a differentially rotating planet would be different from those of a uniformly rotating planet. The differential rotation is chosen to have angular velocity as a function of cylindrical-radial distance only ($\partial \Omega / \partial z =0$). This consideration is mainly motivated by the fact that such a velocity profile does not violate the barotropic assumption, and a full solution to the Euler equation can be obtained without further complications. (Of course, there is no reason to suppose that the planets obey this precisely.) The full solution of the Euler equation can then be compared to that obtained from the thermal wind equation. 

\begin{figure}[h!]
 \centering
     \includegraphics[width=\textwidth]{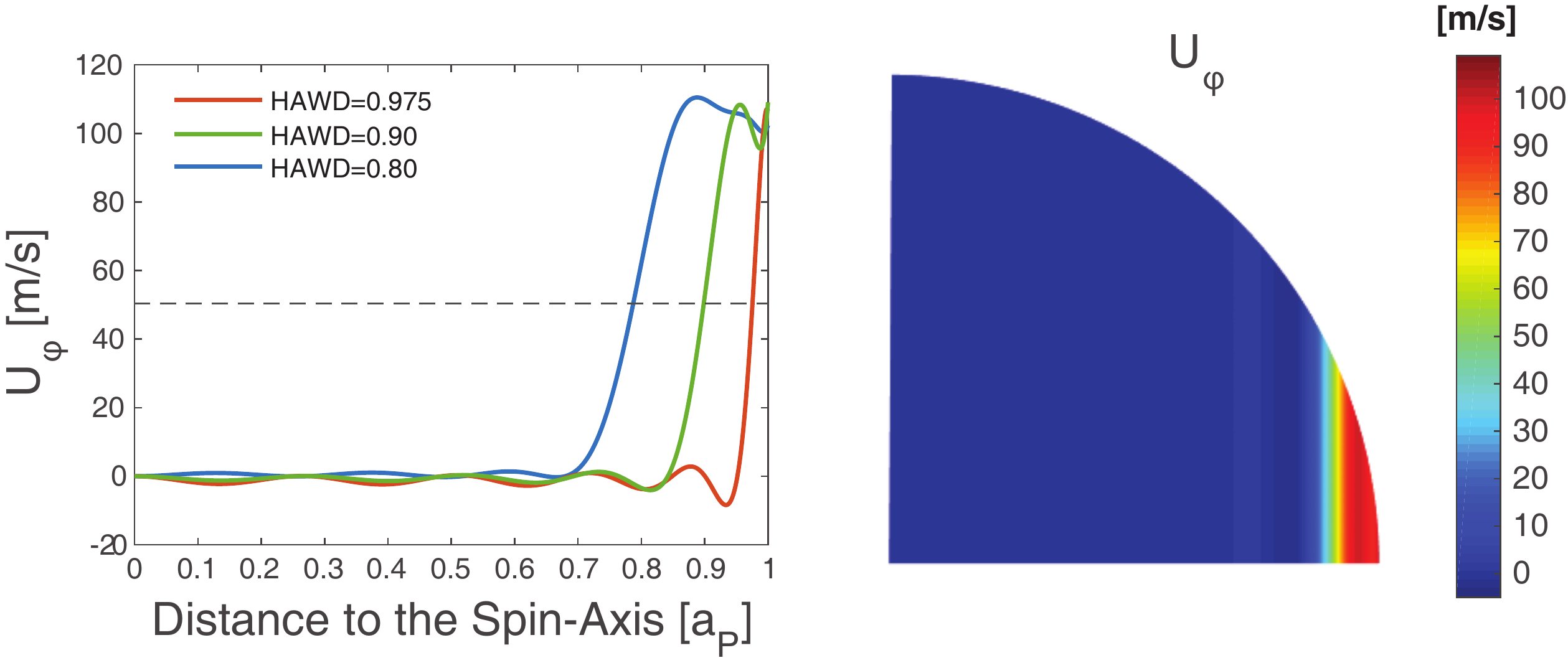}
 \caption{Zonal wind profiles considered in this study. All the zonal winds are assumed to be constant along the direction parallel to the spin-axis. Different wind profiles are characterized by the different half-amplitude-width (HAWD), which is defined as the fractional cylindrical radius at which the amplitude of the zonal wind equals half of the peak amplitude.}
\end{figure}

Fig. 3 shows a series of zonal wind profiles adopted in this study. All the zonal wind profiles feature a prominant band of equatorial super rotation. Different wind profiles are characterized by the different half-amplitude-width (HAWD), which is defined as the cylindrical radius at which the amplitude of the zonal wind equals half of the peak amplitude. To ensure convergence, the generalized centrifugal potential (\ref{eqn:Q1}) is approximated by a polynomial expansion of cylindrical radius $s$ and truncated at degree 24 following \citet[][]{Hubbard1982, Kaspi2016, WH2016, Galanti2017}. The density perturbations and gravity moments associated with this wind profile are then calculated using four different approaches: the Bessel method (full solution to the Euler equation), the CMS method (full solution to the Euler equation), the thermal wind equation with spherical background state, and the thermal wind equation with non-spherical background state.


\subsection{Euler Equation Solution from the Bessel Method and the CMS Method}

With differential rotation on cylinders and a polytrope of index unity, the divergence of the Euler equation (\ref{eqn:Euler}) now reads
\begin{equation}
\nabla^2 \rho +\frac{2\pi G}{K} \rho = -\nabla^2 (\frac{Q}{2K}),
\label{eqn:EulerZonalFlow}
\end{equation}
where
$Q$ is the generalized centrifugal potential (\ref{eqn:Q1}). 

For the Bessel method, the solution to this equation takes the same functional form as (\ref{eqn:RhoPoly1}). The difference is that the specific solution $\rho_P$ is now a function of cylindrical radius $s$ rather than a constant. The specific solution $\rho_P(s)$ \citep[e.g.][]{Hubbard1999} for the wind profile we are considering can be obtained via numerically integrating equation (\ref{eqn:EulerZonalFlow}) with the inner boundary condition $\rho_P(0)=\Omega_0^2/2 \pi G$. This inner boundary condition is only valid when the coefficient of the $s^2$ term in the polynomial expansion of $Q=\sum_{i=1}^{12} B_{2i} s^{2i}$ is set to $1/2\Omega_0^2$. After obtaining $\rho_P(s)$, we can solve for the internal density distribution and gravity moments $J_n$ associated with the differential rotation using the same non-perturbative approach. 

For the CMS method, one only needs to replace the centrifugal potential term $Q_0$ in the equipotential surface equations with the generalized centrifugal potential term $Q$. There is no need to obtain a specific solution $\rho_P$ in the CMS method. 

\begin{table}[t!]
\setlength{\tabcolsep}{9pt}
\renewcommand{\arraystretch}{1.2}
\caption{Zonal wind induced $\Delta J_n$ calculated from Bessel method and CMS method for $\text{HAWD}=0.80$.\label{tab2}}
\centering
\begin{tabular}{ c | r r r r} 
\hline
\hline
 [$10^{-6}$] & Bessel Method & CMS (521) & | Bessel - CMS (521) | & Fractional Difference \\
\hline
$\Delta J_{2}$ & $46.045$  &  $46.290$ & $2.45 \times 10^{-1}$ & $3.30 \times 10^{-4}$\\ 
$\Delta J_{4}$ &  $-16.861$  & $-16.929$ & $6.71 \times 10^{-2}$ & $7.22 \times 10^{-4}$\\ 
$\Delta J_{6}$ &  $4.856$  & $4.873$ & $1.62 \times 10^{-2}$ & $1.25 \times 10^{-3}$\\ 
$\Delta J_{8}$ &  $-0.510$  & $-0.508$ & $2.65 \times 10^{-3}$ & $4.52 \times 10^{-5}$\\ 
\hline
$\Delta J_{10}$ & $-0.394$ & $-0.398$ & $4.35 \times 10^{-3}$ & $1.84 \times 10^{-2}$ \\
$\Delta J_{12}$ & $-0.203$ & $-0.204$ & $1.01 \times 10^{-3}$ & $5.24 \times 10^{-3}$ \\
$\Delta J_{14}$ & $0.0202$ & $0.0211$ & $8.67 \times 10^{-4}$ & $4 \times 10^{-2}$ \\
$\Delta J_{16}$ & $-0.0575$ & $-0.0582$ & $6.41 \times 10^{-4}$ & $1.11 \times 10^{-2}$ \\
$\Delta J_{18}$ & $0.0146$ & $0.0146$ & $6.35 \times 10^{-5}$ & $4.33 \times 10^{-3}$ \\
$\Delta J_{20}$ & $0.0105$ & $0.0108$ & $2.75 \times 10^{-4}$ & $2.61 \times 10^{-2}$ \\ 
\hline
\end{tabular}
\end{table}

\begin{table}[t!]
\setlength{\tabcolsep}{9pt}
\renewcommand{\arraystretch}{1.2}
\caption{Zonal wind induced $\Delta J_n$ calculated from Bessel method and CMS method for $\text{HAWD}=0.90$.\label{tab3}}
\centering
\begin{tabular}{ c | r r r r} 
\hline
\hline
 [$10^{-6}$] & Bessel Method & CMS (521) & | Bessel - CMS (521) | & Fractional Difference \\
\hline
$\Delta J_{2}$ & $5.835$  &  $5.882$ & $4.69 \times 10^{-2}$ & $3.17 \times 10^{-4}$\\ 
$\Delta J_{4}$ &  $-4.662$  & $-4.679$ & $1.75 \times 10^{-2}$ & $6.46 \times 10^{-4}$\\ 
$\Delta J_{6}$ &  $2.495$  & $2.512$ & $1.74 \times 10^{-2}$ & $1.38 \times 10^{-3}$\\ 
$\Delta J_{8}$ &  $-1.147$  & $-1.157$ & $1.04 \times 10^{-2}$ & $3.96 \times 10^{-3}$\\ 
\hline
$\Delta J_{10}$ & $-0.397$ & $-0.400$ & $3.40 \times 10^{-3}$ & $6.42 \times 10^{-3}$ \\
$\Delta J_{12}$ & $-0.0439$ & $-0.0431$ & $7.50 \times 10^{-4}$ & $1.22 \times 10^{-2}$ \\
$\Delta J_{14}$ & $-0.0695$ & $-0.0716$ & $2.06 \times 10^{-3}$ & $3.03 \times 10^{-2}$ \\
$\Delta J_{16}$ & $0.0670$ & $0.0686$ & $1.65 \times 10^{-3}$ & $2.47 \times 10^{-2}$ \\
$\Delta J_{18}$ & $-0.0281$ & $-0.0288$ & $7.40 \times 10^{-4}$ & $2.64 \times 10^{-2}$ \\
$\Delta J_{20}$ & $-0.00237$ & $-0.00224$ & $1.36 \times 10^{-4}$ & $5.74 \times 10^{-2}$ \\ 
\hline
\end{tabular}
\end{table}

\begin{table}[t!]
\setlength{\tabcolsep}{9pt}
\renewcommand{\arraystretch}{1.2}
\caption{Zonal wind induced $\Delta J_n$ calculated from Bessel method and CMS method for $\text{HAWD}=0.975$.\label{tab4}}
\centering
\begin{tabular}{ c | r r r r} 
\hline
\hline
 [$10^{-6}$] & Bessel Method & CMS (521) & | Bessel - CMS (521) | & Fractional Difference \\
\hline
$\Delta J_{2}$ & $1.963$  &  $1.929$ & $3.40 \times 10^{-2}$ & $3.11\times 10^{-4}$\\ 
$\Delta J_{4}$ &  $-0.380$  & $-0.343$ & $3.77 \times 10^{-2}$ & $5.47 \times 10^{-4}$\\ 
$\Delta J_{6}$ &  $0.189$  & $0.169$ & $1.69 \times 10^{-2}$ & $3.84 \times 10^{-4}$\\ 
$\Delta J_{8}$ &  $-0.128$  & $-0.121$ & $7.15 \times 10^{-3}$ & $2.06 \times 10^{-3}$\\ 
\hline
$\Delta J_{10}$ & $0.0929$ & $0.0904$ & $3.40 \times 10^{-3}$ & $8.45 \times 10^{-3}$ \\
$\Delta J_{12}$ & $-0.0685$ & $-0.0683$ & $2.45 \times 10^{-4}$ & $2.60 \times 10^{-3}$ \\
$\Delta J_{14}$ & $0.0488$ & $0.0495$ & $7.09 \times 10^{-4}$ & $1.42 \times 10^{-2}$ \\
$\Delta J_{16}$ & $-0.0374$ & $-0.0383$ & $9.14 \times 10^{-3}$ & $2.44 \times 10^{-2}$ \\
$\Delta J_{18}$ & $0.0263$ & $0.0270$ & $6.18 \times 10^{-4}$ & $2.35 \times 10^{-2}$ \\
$\Delta J_{20}$ & $-0.00733$ & $-0.00754$ & $2.11 \times 10^{-4}$ & $2.87 \times 10^{-2}$ \\ 
\hline
\end{tabular}
\end{table}

The left panel of Fig. 4 compares the total gravity moments $J_n$ associated with the zonal flows shown in Fig. 3 to the gravity moments associated with the background uniform rotation, while the right panel of Fig. 4 compares the wind-induced gravity moments only, defined as $\Delta J_n = J_n (\text{Uniform Rotation + Wind}) - J_n (\text{Uniform Rotation})$. The solution obtained from the Bessel method and the CMS method are in very good agreement and appear almost identical in the figure, thus only the solution from the Bessel method is shown here for clarity. Tables 2-4 presented the tabulated comparison between the Bessel method and the CMS method for the wind induced gravity moments. It can be seen from the tables that these two methods agree very well for all three different wind profiles: 1) the absolute differences in $\Delta J_n$ are smaller than $3 \times 10^{-9}$ beyond degree 10, and 2) the fractional differences in $J_n$ are on the order of 3\%, except for Jn with absolute values smaller than $3 \times 10^{-9}$. In contrast, \citet[]{Kaspi2016} reported much larger discrepancies between the CMS method and the Bessel method. In \citet[]{Kaspi2016}, the absolute differences in high-degree $J_n$ are typically on the order of $1 \times 10^{-7}$, and the fractional differences in high-degree $J_n$ are typically on the order of 30\%. We are confident in the good agreement we obtained between the CMS method and the Bessel method, given that \citet[]{WH2016} have independently shown the very good agreement between these two methods for the uniform rotation case.

It should be emphasized here that in both the Bessel method and the CMS method, we are solving the full Euler equation to get the total gravity moments for the case with zonal flows, rather than solving for the perturbations $\Delta J_n$ only. This is fundamentally different from the approach taken by \citet[]{Kong2014, Kong2015}, in which the wind induced gravity is treated as a perturbation. One aspect of the full solution is that the outer boundary shape of the planet gets further changed when zonal winds are included. 

It can be seen from the left panel of Fig. 4 that the gravity moments associated with the zonal flows shown in Fig. 3 exceed those from the background rotation by more than 100\% starting around degree 10. For the low-degree gravitational moments, the wind induced contribution increases as the depth of the wind increases (see Fig. 4 \& Tab. 2 - 4). For the high-degree gravitational moments, however, the wind induced contributions are similar for the three different wind depth considered here (see Fig. 4 \& Tab. 2 - 4). This feature should caution about the attempt to estimate the wind-induced perturbations to the low-degree gravity moments based on future measurements of the high-degree gravity moments. 

\begin{figure}[h!]
 \centering
     \includegraphics[width=\textwidth]{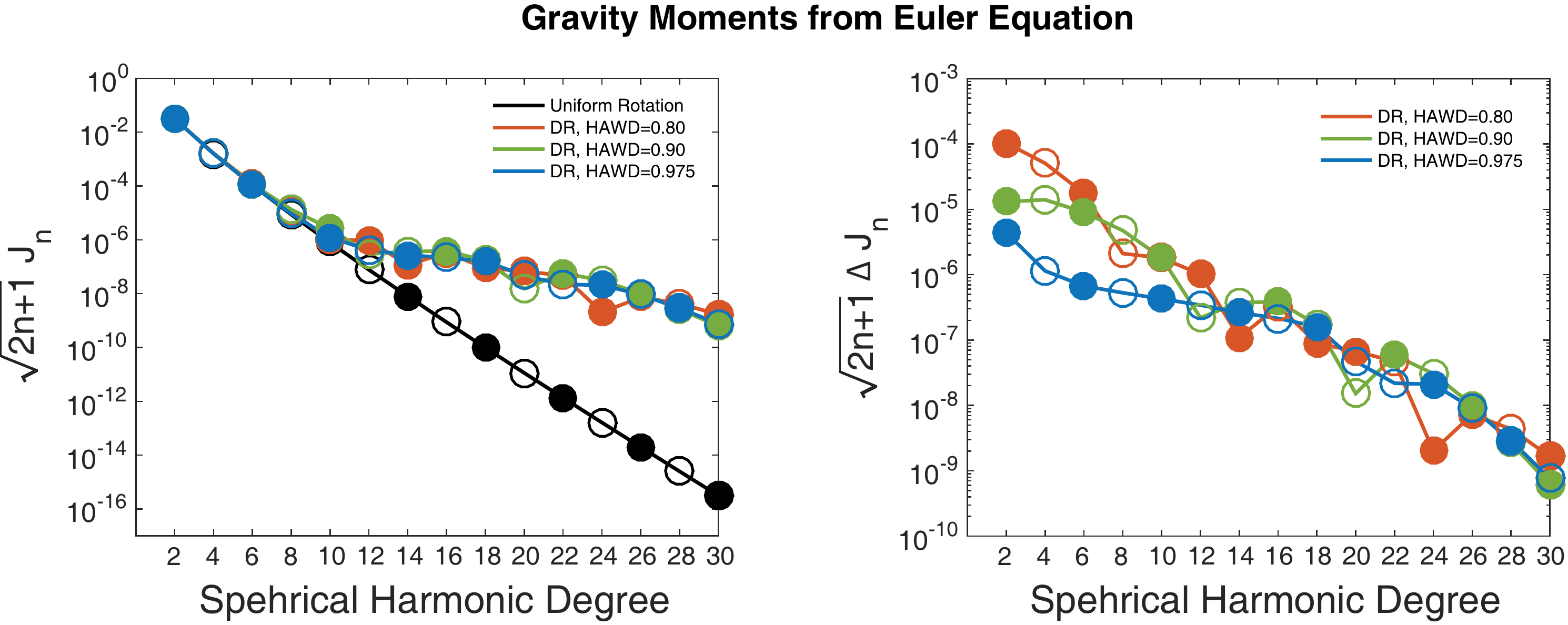}
 \caption{Gravity moments associated with deep equatorial zonal flows calculated from the Euler equation. $J_n$ are shown in the left panel, while $\Delta J_n = J_n (\text{Differential Rotation}) - J_n (\text{Uniform Rotation})$ are shown in the right panel. For $J_n$ and $\Delta J_n$, filled (open) circles represent positive (negative) values.}
\end{figure}

 \subsection{Thermal Wind Equation with Spherical Background Density and Gravity}

\begin{figure}[h!]
 \centering
     \includegraphics[width=0.75\textwidth]{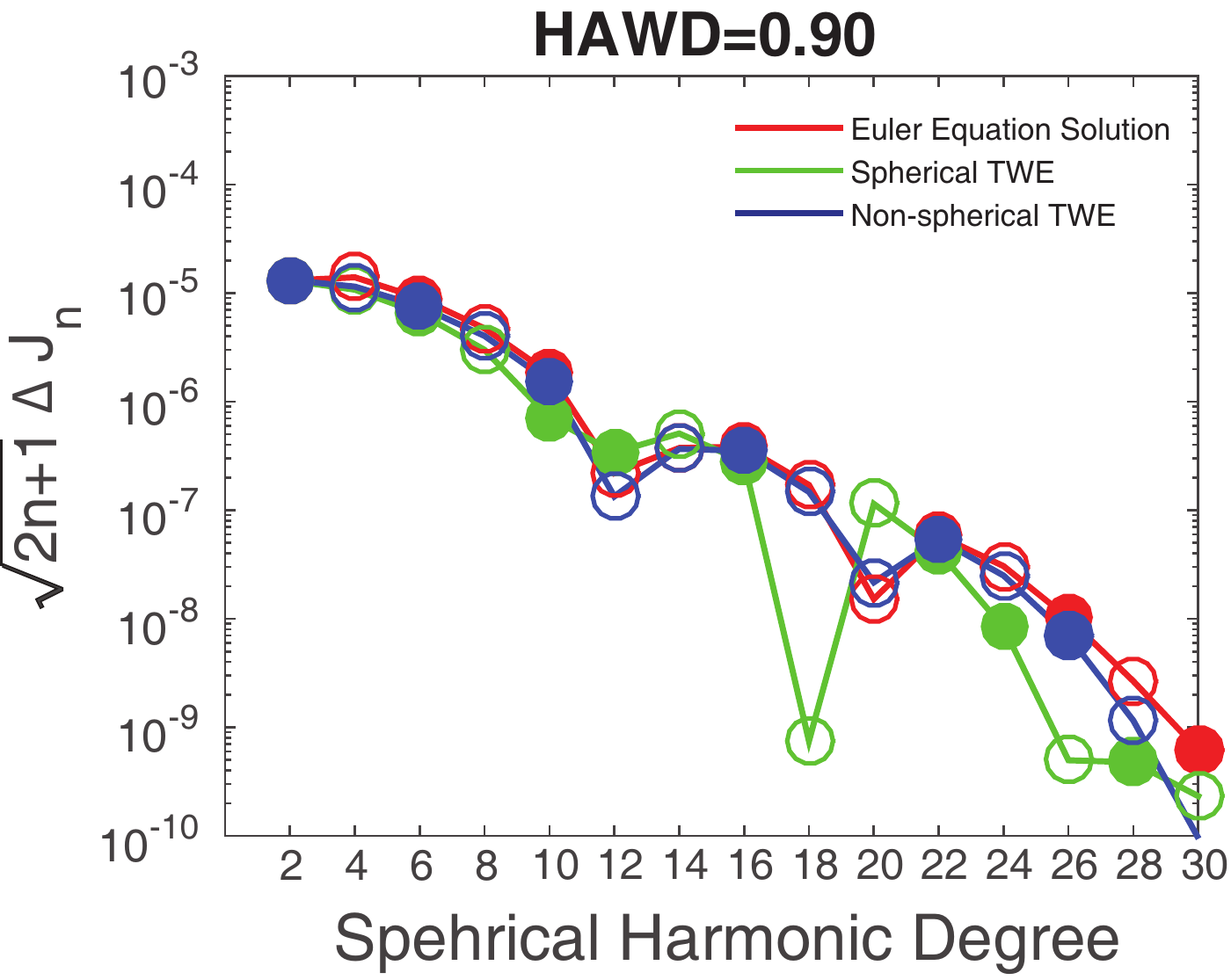}
 \caption{Gravity moments associated with deep equatorial zonal flows calculated from the Euler equation compared to those calculated from the spherical thermal wind equation and the non-spherical thermal wind equation for HAWD=0.90. Only $\Delta J_n$ are shown here, and filled (open) circles represent positive (negative) values.}
\end{figure}

\begin{figure}[h!]
 \centering
     \includegraphics[width=\textwidth]{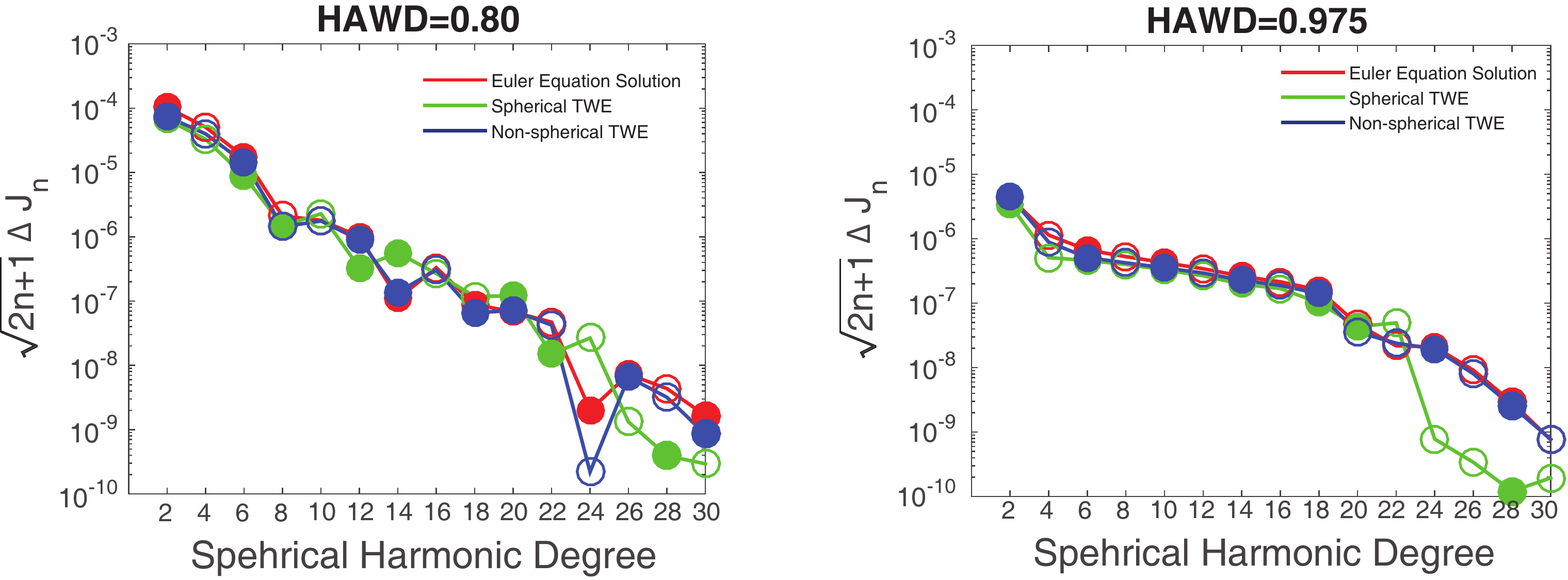}
 \caption{Gravity moments associated with deep equatorial zonal flows calculated from the Euler equation compared to those calculated from the spherical thermal wind equation and the non-spherical thermal wind equation for HAWD=0.80 \& HAWD=0.975. Only $\Delta J_n$ are shown here, and filled (open) circles represent positive (negative) values.}
\end{figure}

We now proceed to solve for the gravity moments and density perturbations associated with the same zonal flows using the thermal wind equation. We first consider the thermal wind equation under the simplification that reduces both the background density and the background effective gravity to a spherically symmetric state. This is the simplification adopted in many published calculations of the zonal flow gravity using the thermal wind equation \citep[e.g.][]{Kaspi2010, Liu2013}. The thermal wind equation now reads
\begin{equation}
2|\Omega_0| \frac{\partial [\rho_0 (r) \mathbf{u'}]}{\partial z}=- \frac{1}{r} \frac{\partial \rho'}{\partial \theta} \mathbf{e_\theta} \times (-|g_0(r)| \mathbf{e_r}).
\label{eqn:sTWE}
\end{equation}
The spherically symmetric background density and background effective gravity corresponding to a planet with the same mass and the same polytropic index unity equation of state can be obtained analytically.

One motivation for adopting a spherically symmetric background state in the thermal wind equation to calculate the wind induced gravity moments is the uniqueness of $J_n$ despite the non-uniqueness of wind induced density perturbations. With spherically symmetric background state, the TWE yields the gradient of the density perturbations $\nabla \rho'$ along the $\theta$ direction ${\partial \rho'}/{\partial \theta}$. To get the density perturbations $\rho'(r, \theta)$ with spherically symmetric background state, one would integrate ${\partial \rho'}/{\partial \theta}$ along the $\theta$ direction
\begin{equation}
\rho'(r,\theta)=\int_{\theta'=0}^{\theta'=\theta} \frac{\partial \rho'}{\partial \theta} rd\mathbf{\theta'}+\rho'_c(r),
\label{eqn:InteTWE}
\end{equation}
here $\rho'_c(r)$ is a ``constant of integration" which is a function of $r$ only. This "constant of integration" resulting from spherical thermal wind equation makes zero contribution to the gravity moments with $n \ge 1$, since 
\begin{equation}
\int_{\mathbf{R}^3} \rho'_c(r) r^n P_n(cos\theta) d^3\mathbf{r}=0, \qquad{} n \ge 1.
\label{eqn:ConsInt}
\end{equation}

The gravity moments associated with the zonal flows shown in Fig. 3 calculated from the thermal wind equation with spherically symmetric background density and spherically symmetric background gravity is compared to those calculated from the Euler equation in Fig. 5 \& 6 and in Tab. 5. Only $\Delta J_n$ are shown in Fig. 5 \& 6 and in Tab. 5. It can be seen that 1) overall, the solution from the spherical TWE is in order-of-magnitude agreement with the full solution; 2) however, some of the individual high-degree gravity moments calculated from this simplified thermal wind equation can be wrong by more than 100\% and can take the wrong sign. This is consistently the case for the series of zonal wind profiles adopted in this study. 

\begin{table}[t!]
\setlength{\tabcolsep}{9pt}
\renewcommand{\arraystretch}{1.2}
\caption{Zonal wind induced $\Delta J_n$ for \text{HAWD}=0.90 calculated from four methods.\label{tab5}}
\centering
\begin{tabular}{ c | r r r r} 
\hline
\hline
 [$10^{-6}$] & Bessel & CMS & Spherical & Non-spherical \\
 & Method & Method &TWE & TWE \\
\hline
$\Delta J_{2}$ & $5.835$  &  $5.882$ & $5.836$ & $5.897$\\ 
$\Delta J_{4}$ &  $-4.662$  & $-4.679$ & $-3.586$ & $-3.826$\\ 
$\Delta J_{6}$ &  $2.495$  & $2.512$ & $1.810$ & $2.098$\\ 
$\Delta J_{8}$ &  $-1.147$  & $-1.157$ & $-0.728$ & $-0.978$\\ 
\hline
$\Delta J_{10}$ & $0.397$ & $0.400$ & $0.155$ & $0.333$ \\
$\Delta J_{12}$ & $\mathbf{-0.0439}$ & $\mathbf{-0.0431}$ & $\mathbf{0.0682}$ & $\mathbf{-0.0268}$\\
$\Delta J_{14}$ & $-0.0695$ & $-0.0716$ & $-0.0945$ & $-0.0681$ \\
$\Delta J_{16}$ & $0.0670$ & $0.0686$ & $0.0486$ & $0.0614$ \\
$\Delta J_{18}$ & $\mathbf{-0.0281}$ & $\mathbf{-0.0288}$ & $\mathbf{-0.0001}$ & $\mathbf{-0.0242}$ \\
$\Delta J_{20}$ & $\mathbf{-0.00237}$ & $\mathbf{-0.00224}$ &$\mathbf{-0.01786}$ & $\mathbf{-0.00338}$\\  
\hline
\end{tabular}
\end{table}

\subsection{Thermal Wind Equation with Non-spherical Background Density and Effective Gravity}

We now proceed to calculate the density perturbation and gravity moments associated with zonal flows using the generic thermal wind equation with non-spherical background density distribution and non-spherical background effective gravity
\begin{equation}
(2\mathbf{\Omega_0} \cdot \nabla)(\rho_0 \mathbf{U})=-\nabla \rho' \times \mathbf{g_{eff}}.
\label{eqn:eTWE2}
\end{equation} 

For a polytrope of index of unity, the background effective gravity $\mathbf{g_{eff}}$ is simply
\begin{equation}
\mathbf{g_{eff}}=-\nabla U_0=\frac{\nabla P_0}{\rho_0}=2K\nabla \rho_0,
\label{eqn:EffGravRho} 
\end{equation}
which can be easily calculated since $\nabla \rho_0$ is entirely determined by the coefficients $A_n$.


With non-spherical effective gravity, the TWE now yields the gradient of the density perturbations $\nabla \rho'$ along the tangent of equipotential surfaces in the meridional plane instead of the gradient of the density perturbations along the $\theta$ direction. To get the density perturbations $\rho'(r, \theta)$ with non-spherical effective gravity, one would need to integrate $\nabla \rho'$ along the tangent of equipotential surfaces in the meridional plane
\begin{equation}
\rho'(\xi,l)=\int_{l'=0}^{l'=l} \nabla \rho' \cdot d\mathbf{l'}+\rho'_c(\xi),
\label{eqn:InteTWE}
\end{equation}
here $\xi$ is measured along the direction perpendicular to the equal-potential surface, $l$ and $dl'$ are the meridional arc length measured on the equal-potential surface, and $\rho'_c(\xi)$ is a ``constant of integration" which is a function of $\xi$ instead of $r$. This "constant of integration" resulting from non-spherical thermal wind equation makes a non-zero contribution to the gravity moments. Since $\nabla \rho'_c(\xi) \times \mathbf{g_{eff}}=0$, this ``constant of integration" has zero contribution to the generic thermal wind equation. It is clear then that TWE itself cannot supply the "constant of integration". However, since $\rho'_c(\xi)$ is only a function of $\xi$, the gravity moments associated with this ``constant of integration" should be proportional to the background $J_n$, and the pre-factor should be on the order of the ratio of the wind-induced mass anomaly to the total mass of the planet which is a very small quantity. Since the wind-induced $J_n$ are orders of magnitude larger than the background $J_n$ beyond $n=10$, the correction to $J_n$ by $\rho'_c(\xi)$ must be negligible beyond $n=10$. Thus we set $\rho'_c(\xi)$ to zero in all our non-spherical TWE calculations.


It can be seen from Fig. 5 \& 6 and Tab. 5 that for $n \ge 10$, $\Delta J_n$ associated with zonal flows calculated from the thermal wind equation with non-spherical background state is much closer to the full solution than those calculated from the thermal wind equation with spherical background state. All $\Delta J_n$ greater than $3 \times 10^{-9}$ now take the correct sign, and the amplitude difference is now within 50\% for individual $\Delta J_n$.

\begin{figure}[h!]
 \centering
     \includegraphics[width=\textwidth]{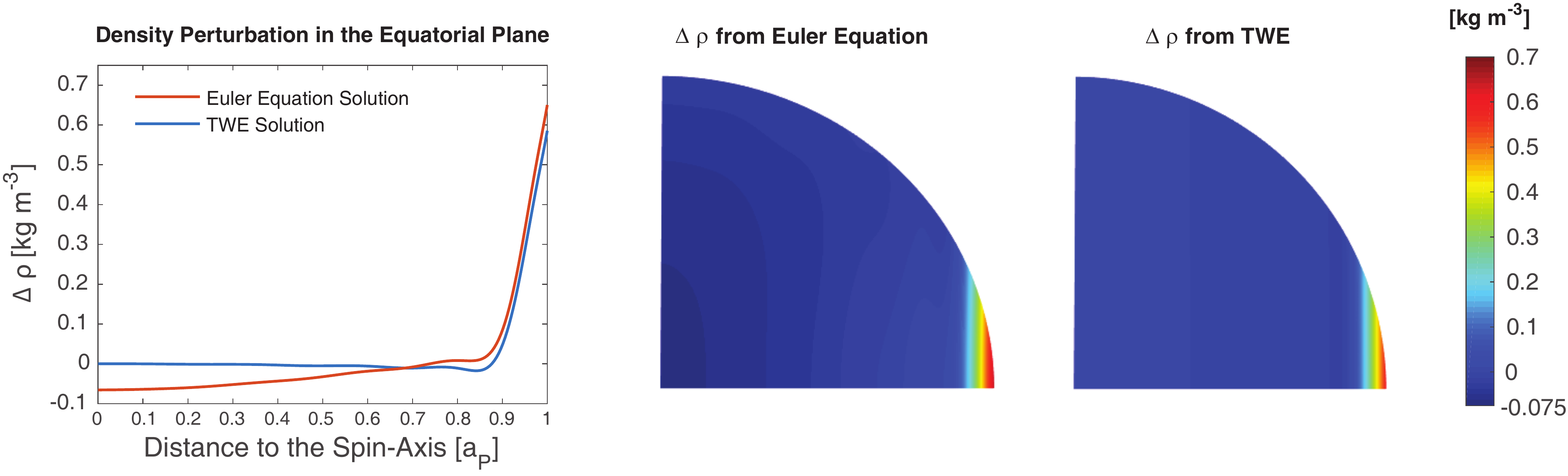}
 \caption{Density perturbations associated with deep zonal flows (the case with \text{HAWD}=$0.90$ in Fig. 3) calculated from the Euler equation and the thermal wind equation. The leftmost panel shows the density perturbations in the equatorial plane while the right two panels show the meridional cut. It can be seen that the thermal wind equation captures the local wind-induced density perturbations but misses the non-local large-scale density perturbations.}
\end{figure}

\section{What the Thermal Wind Equation Misses: Global Shape Change, Non-local and Irrotational Density Contributions}

As discussed in section 3.3, density perturbations that contribute irrotationally to the Euler equation and that have a non-local origin must necessarily exist. Fig. 7 compares the density perturbations associated with deep zonal flows (the case with \text{HAWD}=$0.90$ in Fig. 3) calculated from the Euler equation and the thermal wind equation. The density perturbations calculated from two version of the thermal equation are both localized and are visually similar, thus only the density perturbations calculated from the thermal wind equation with spherical background state is shown for clarity. It can be seen from Fig. 7 that the thermal wind equation captures the local density perturbations directly associated with the local zonal flows but misses the large-scale density perturbations (with a dominant degree-2 structure) associated with the global shape change of the planet related to the net positive angular moment of the zonal flows. The zonal wind profiles we are considering all have net positive total angular moment. As a result the planet shrinks in the polar direction while it expands in the equatorial direction.  

\begin{figure}[h!]
 \centering
     \includegraphics[width=0.5\textwidth]{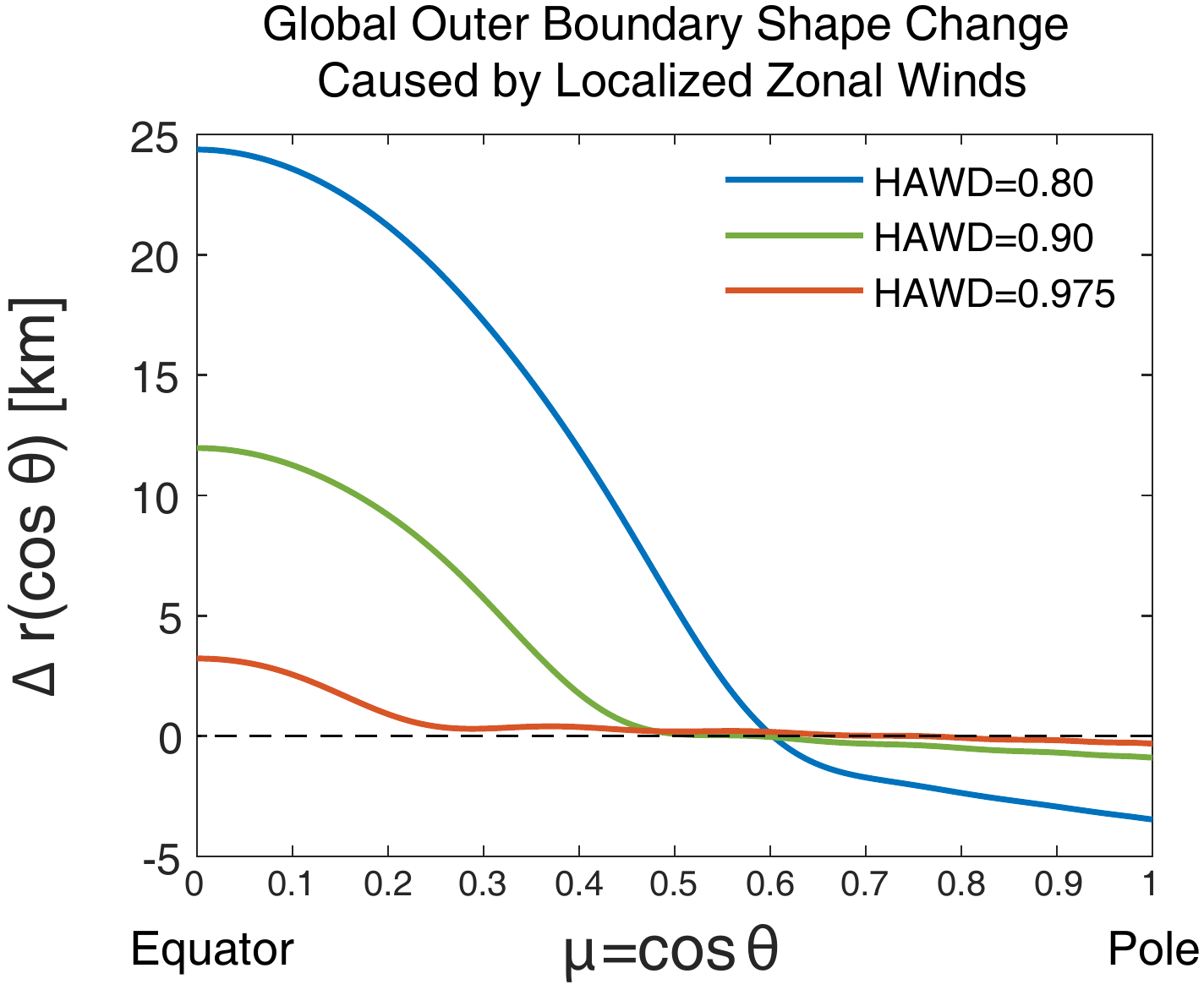}
 \caption{Global outer boundary shape change induced by the localized deep zonal flow shown in Fig. 3. It can be seen that the localized equatorial jet with half amplitude width (HAWD) of 0.80 can induce a change to the outer boundary position by $\sim$ 25 $km$ near the equator and by $\sim$ 4 $km$ near the poles where there are no local zonal flows.}
\end{figure}

Fig. 8 shows the outer boundary shape change of the entire planet caused by the localized zonal flows shown in Fig. 3. It can be seen that the equatorial radius increased by $\sim$ 25 $km$, while the polar radius decreased by $\sim$ 4 $km$ for an equatorial super rotation with half amplitude width at 0.80 $R_J$. Small-scale shape changes spatially correlated with the local zonal flows are also evident Fig. 8. 

\citet[]{ZKS2015} proposed a gravitational correction to the thermal wind equation by including the gravitational anomalies from the local density perturbation required by the TWE: the curl of the $\rho_0 \nabla V_{g'}$ term. We performed an independent calculation of the gravitational anomaly correction as proposed in \citet[]{ZKS2015}, and find that for the high-degree gravity moments beyond $n=12$, the gravitational anomaly term can only provide corrections on the order of $1\%$. This is consistent with the independent analysis of \citet[]{Galanti2017}.


A cautionary note about the wind-induced low-degree gravity moments: they are insignificant corrections to the background low-degree gravity moments (e.g 0.3\% correction to $J_2$ and 3\% to $J_4$ for the deepest wind profile considered here with $\text{HAWD}=0.80 R_J$; 0.01\% correction to $J_2$ and 0.07\% correction to $J_4$ for $\text{HAWD}=0.975 R_J$), and can be easily offset by uncertainties in the background state (such as uncertainties in our knowledege about the equation of state, thermal state, heavy element distribution, etc.)

\section{Summary and Discussion}

In this paper, we present a critical examination of the applicability of the thermal wind equation under anelastic assumption to calculate the gravity moments associated with deep zonal flows of giant planets. We first derive the thermal wind equation from the Euler equation and show that the thermal wind equation is a good approximation to the local dynamics when the Rossby number of the zonal flow measured in the co-rotation frame is much smaller than one. It is also pointed out that the thermal wind equation is a local treatment even when the background effective gravity is used, while the full problem is non-local. 

We then solve the full Euler equation for a rotating self-gravitating planet with polytrope of index unity. The Bessel method \citep[]{Hubbard1975, Wisdom1996, Hubbard1999} and the Concentric Maclaurin Spheroid method \citep[]{Hubbard2013} are employed. We first solve for the shape, density distribution, and the gravity moments of a uniformly rotating planet. It is shown that the outer boundary shape has a significant deviation, on the second order, from an exact ellipsoid of revolution. The impact of the assumption that the outer boundary shape is an exact ellipsoid of revolution adopted in some studies of this problem \citep[e.g.][]{Kong2013, Kong2015} on the solutions of gravity moments requires further investigation.

For Jupiter-like zonal flows but confined to the equatorial region and assumed to be constant on cylinders (e.g. Fig. 3), the associated density perturbations and gravity moments are then calculated from four different methods: the Bessel method, the CMS method, the thermal wind equation with spherical background state, and the thermal wind equation with non-spherical background state. The full solutions to the Euler equation obtained from the Bessel method and the CMS method show excellent agreement.

Concerning the applicability of the thermal wind equation, we find that 1) overall, the solution from the spherical TWE is in order-of-magnitude agreement with the full solution; 2) a few individual high-degree gravity moments calculated from the spherical thermal wind equation can be wrong by 100\% and can take the wrong sign; 3) the individual high-degree gravity moments calculated from the thermal wind equation with non-spherical background density and non-spherical effective gravity are a good approximation to the full solution, the difference is within 50\%; 4) for low-degree gravity moments associated with zonal flows, global shape change to the planet caused by the net angular moments of the zonal flows is important. This global shape change is missed in the thermal wind equation as well as in the thermal-gravitational wind equation \citep[]{ZKS2015}. However, the wind-induced low-degree gravity moments may not be a concern since they are most likely indiscernible from uncertainties in the background state.

For baroclinic zonal winds with velocity variations along the direction of spin-axis, we don't know how to solve the full Euler equation to obtain the density perturbation, shape change, and gravity moments. For baroclinic winds, the equipotential surface and equal-density surface are mis-aligned in general. Furthermore, the generalized centrifugal potential $Q$ (e.g. equations 12 - 13) cannot be defined for a baroclinic flow. To be more specific, $(\mathbf{u} \cdot \nabla) \mathbf{u}$ of a baroclinic flow cannot be written as a gradient of a scalar potential. The current technique to solve the Euler equation, such as the Bessel method and the CMS method, iteratively find the solution via requiring the equal-density surface being an equipotential surface, thus cannot be straightforwardly generalized to deal with baroclinic flows. Non-perturbative methods are yet to be developed to solve the full equation for baroclinic zonal flows. The thermal wind equation, on the other hand, remains a valid approximation for low-Rossby flows. Furthermore, the thermal wind equation, with spherical or non-spherical background state, requires significantly less computational resources than the full Euler equation.

A clear message from this study to the analysis of the gravity measurements from \textit{Juno} and \textit{Cassini} is that to calculate the wind-induced gravity moments using the thermal wind equation, taking the non-spherical nature of the background density and effective gravity into account would yield much more accurate representation of the full solution.

\acknowledgments
This work has been supported by NASA's Juno mission. All relevant data are listed in the tables and in the manuscript. 
We thank the two reviewer for their constructive comments.

\listofchanges

\end{document}